\documentclass[fleqn,10pt]{wlscirep}
\title{Strong light illumination on gain-switched semiconductor lasers helps the eavesdropper in practical quantum key distribution systems}

\author[1,*]{Yang-yang Fei}
\author[1]{Xiang-dong Meng}
\author[1]{Ming Gao}
\author[1]{Yi Yang}
\author[1]{Hong Wang}
\author[1,2]{Zhi Ma}
\affil[1]{State Key Laboratory of Mathematical Engineering and Advanced Computing, Zhengzhou, Henan, 450001, China}

\affil[2]{CAS Center for Excellence and Synergetic Innovation Center in Quantum Information and Quantum Physics, University of Science and Technology of China, Hefei, Anhui, 230026, China}

\affil[*]{fei\_yy@foxmail.com}

\begin{abstract}
The temperature of the semiconductor diode increases under strong light illumination whether thermoelectric cooler is installed or not, which changes the output wavelength of the laser (Lee M. S. et al., 2017). However, other characteristics also vary as temperature increases. These variations may help the eavesdropper in practical quantum key distribution systems. We study the effects of temperature increase on gain-switched semiconductor lasers by simulating temperature dependent rate equations. The results show that temperature increase may cause large intensity fluctuation, decrease the output intensity and lead the signal state and decoy state distinguishable. We also propose a modified photon number splitting attack by exploiting the effects of temperature increase. Countermeasures are also proposed.
\end{abstract}
\begin{document}

\flushbottom
\maketitle
%
%
\noindent $\textbf{keywords:}$ quantum key distribution, gain-switched semiconductor laser, rate equation, strong light illumination.\par
\noindent $\textbf{PACS number(s):}$ 03.67.Dd, 03.67.Hk
\thispagestyle{empty}

\section{Introduction}
Quantum key distribution (QKD) can provide unconditional security to distribute key between two remote parts with perfect devices\cite{gisin2002quantum,lutkenhaus2009focus,scarani2009security}. However, practical devices always deviate from the models in security proofs. These deviations reduce the secure key and  many attacks on imperfect devices have been proposed to steal information about the final key\cite{huttner1995quantum, makarov2005faked, makarov2006effects, qi2007time, zhao2008quantum, gisin2006trojan, jain2014trojan, weier2011quantum, xu2010experimental, lamas2007breaking, wang2013effect,lydersen2010hacking,li2011,sun2011passive,sun2012partially,tang2013source,sun2014hacking,fei2015}. Thus, study on the imperfections of practical devices is extremely important to the security of QKD systems.\par
The sources of practical QKD systems also suffer from some imperfections. Attenuated laser pulses are always used as the "single photon source" in practical QKD systems due to technique limitation. However, the attenuated pulses may contain more than one photons, which can be attacked by photon number splitting (PNS) attack\cite{huttner1995quantum}. To against PNS attack on weak coherent source, decoy state method is proposed \cite{hwang2003quantum,lo2005decoy,wang2005beating}. In the "weak + vacuum" decoy state method\cite{ma2005practical}, a sender (Alice) needs to transmit pluses of three intensities (signal state, decoy state and vacuum state) to a receiver (Bob). Usually, the intensity of signal state is higher than that of the decoy state.\par
Gain-switched semiconductor laser is widely used in practical QKD systems as the transmitter, because it simplifies the construction of the source and sends phase randomized pulses \cite{2014Evaluation}. The phase randomness of the pulses is assumed in most security proofs. A gain-switched semiconductor laser is driven from the initial carrier density, which is below the threshold, by a strong AC injection current pulse with the level of $J_{AC}$ for each photon pulse generation in QKD systems.
The initial carrier density is determined by a DC bias current, $J_{DC}$. Short pulses are generated by injection of short current pulses into a semiconductor laser. The shape of the output photon pulses are directly defined by $J_{AC}$ and $J_{DC}$. Actually, the carrier density and photon density vary fast and acutely in the gain-switched operation. A rate equation description is used to efficiently and accurately simulate the  performance of gain-switched semiconductor lasers with proper choice of model parameters. The output performance of a semiconductor laser is sensitive to temperature variation\cite{Dousmanis1964TEMPERATURE}. So most semiconductor lasers work with thermoelectric coolers to keep the temperature of semiconductor diode stable.

In Ref. [\citen{Lee2017Free}], Lee $et\ al.$ experimentally show that the temperatures of semiconductor diodes can be increased by strong light illumination, even when the lasers are installed with thermoelectric coolers. Due to the limit power of the thermoelectric cooler, the temperature of the diode increases as long as the illumination is strong enough. Temperature increase leads to output wavelength variation which makes pulses from different lasers distinguishable\cite{Lee2017Free}. However, the wavelength variation do not impact the security of QKD systems with only one laser, such as phase encoding systems. Besides, other characteristics of the semiconductor lasers also change with temperature \cite{Byrne1989A,Dousmanis1964TEMPERATURE}. The variations of other characteristics also deviate the behavior of semiconductor lasers from ideal models in security proofs. And the eavesdropper may exploit these variations to steal information about the final key. So it is urgently needed to study the effects of temperature increase on important characteristics of gain-switched semiconductor lasers.\par
In the following, we mainly focus on the effects of temperature increase on gain-switched semiconductor diodes in QKD systems. This article is constructed as follows: we briefly introduce the temperature dependent single mode rate equation of semiconductor lasers and the model parameters for a gain-switched semiconductor laser used in this paper in Sec. II. Then numerical simulation of the efficient and accurate single mode temperature dependent rate equation is performed in Sec. III. The effects of temperature increase on three characteristics are studied in detail, which include the recovery time of carrier density, the intensity of output pulses and the time interval between signal state and decoy state pulses. 
In Sec. IV, we propose a modified PNS attack strategy which exploits the effects of temperature increase. We also give out the hacking strategy using side channels in time dimension. Finally, we conclude the paper and discuss countermeasures in Sec. V.

\section{Temperature dependent single mode rate equation of semiconductor laser}
Temperature variations effect the output characteristics of semiconductor lasers significantly. Recent research shows the eavesdropper can elevate the working temperature of a semiconductor laser diode by strong light illumination even when thermoelectric cooler is installed. So the effects of temperature increase on the output performance of semiconductor lasers should be taken into consideration in practical QKD systems.\par
The dynamics of semiconductor laser at different temperatures can be described efficiently and accurately with the following single mode rate equations covering temperature\cite{Byrne1989A,SpencerHIGH,Nakata2017Intensity}
\begin{center}
  \begin{equation}
    \frac{dN(t)}{dt}=\frac{J(t)}{qd}-\frac{N(t)}{\tau_n(T)}-g_0(T)[N(t)-N_0(T)]S(t),
   \end{equation}
\end{center}

\begin{center}
  \begin{equation}
    \frac{dS(t)}{dt}=\Gamma g_0(T)[N(t)-N_0(T)]S(t)-\frac{S(t)}{\tau_p} + \frac{\Gamma \beta N(t)}{\tau_n(T)},
   \end{equation}
\end{center}
where $t$ represents time, $T$ is temperature, $q$ is electrical charge, $N(t)$ is the time-variation carrier density and $S(t)$ represents the time-variation photon density. $J(t)$ is the overall injection current and $J(t)=J_{AC}(t)+J_{DC}$. Others are material parameters of semiconductor lasers, which are explained in detail in Table. \ref{tab:example}. Note that $N_0(T)$, $\tau_n(T)$ and $g_0(T)$ are three main temperature related parameters and others can be treated as constants, which are temperature independent.\par

Experiment observations give the following relationship between the threshold current density and the the temperature $J_{th}(T) = J_c exp({T}/{T_0})$\cite{Tucker1984High},
where $T_0$ is the characteristic temperature of the diode and $J_c$ is the current density constant. Therefore we have
\begin{center}
  \begin{equation}
    J_{th}(T+\Delta T) = J_c exp(\frac{T+\Delta T}{T_0})=J_{th}(T)exp(\frac{\Delta T}{T_{0}}).
   \end{equation}
\end{center}
By solving Eqs. (1) and (2), the threshold current density of the diode can also be approximately given out by
\begin{center}
  \begin{equation}
    J_{th}(T)  \approx \frac{q d}{\tau_n(T)}N_{th}(T)=\frac{qd}{\tau_n(T)}[-\frac{1}{g_{0}(T)\Gamma \tau_p} +N_0(T)],
   \end{equation}
\end{center}
where $N_{th}(T)$ is the threshold value of carrier density. Here we use the same models as the ones in Ref. [\citen{Byrne1989A}] to describe $g_{0}(T)$ and $N_{0}(T)$, as shown in Eqs. (5) and (6). The models fit the experiment results very well.
\begin{center}
  \begin{equation}
    g_{0}(T) = g_{0c} exp(\frac{-T}{T_{0a}}),
   \end{equation}
\end{center}

\begin{center}
  \begin{equation}
    N_{0}(T) = N_{0c} exp(\frac{T}{T_{0a}}),
   \end{equation}
\end{center}
where $g_{0c}$ is the differential gain coefficient constant, $N_{0c}$ is the transparent carrier density constant and $T_{0a}$ is the characteristic temperature of the active region.
So we have $g_{0}(T+\Delta T) = g_0(T)exp(-\Delta T/T_{0a})$, $N_{0}(T+\Delta T) = N_{0}(T) exp({\Delta T}/{T_{0a}})$ and
\begin{center}
  \begin{equation}
    J_{th}(T+\Delta T)  \approx \frac{qd}{\tau_n(T+\Delta T)}[\frac{1}{g_{0}(T+\Delta T)\Gamma \tau_p} +N_0(T+\Delta T)]=\frac{\tau_n(T)}{\tau_n(T+\Delta T)}J_{th}(T)exp(\frac{\Delta T}{T_{0a}}).
   \end{equation}
\end{center}
By combining Eqs. (3) and (7), we can get
\begin{center}
  \begin{equation}
    {\tau_n(T+\Delta T)}=\tau_n(T) \frac {exp(\frac{\Delta T}{T_{0a}})}{exp(\frac{\Delta T}{T_{0}})}.
   \end{equation}
\end{center}
Until now, all the three temperature dependent parameters can be calculated at different temperatures. And we can perform numeral simulation of the rate equations (Eqs. (1) and (2)) at different temperatures.

\begin{table}[ht]
\centering
\begin{tabular}{ccccc}
\hline
Number & Parameter symbol & Description & Value & Units  \\
\hline
1 & $g_0(T)$ & temperature dependent differential gain coefficient & $2\times10^{-6}$  (at $25^\text{o}C$)\cite{Nakata2017Intensity} & $cm^3\cdot s^{-1}$ \\
2 & $N_0(T)$ & temperature dependent transparent carrier density & $10^{18}$  (at $25^\text{o}C$)\cite{Nakata2017Intensity} & $cm^{-3}$ \\
3 & $\tau_n(T)$ & temperature dependent carrier lifetime & 1.2 (at $25^\text{o}C$)\cite{Nakata2017Intensity} & $ns$ \\
4 & $\tau_p$ & photon lifetime & 5.0 \cite{Nakata2017Intensity}& $ps$ \\
5 & $\beta$ & fraction of spontaneous emission coupled into lasing mode & 0.001 & - \\
6 & $d$ & thickness of active region & 0.1 & $\mu m$ \\
7 & $\Gamma$ & mode confinement factor & 0.5 & - \\
8 & $J_{AC}$ & AC current density injected into active region & $2.4\times 10^4$ & $A\cdot cm^{-2}$ \\
9 & $J_{DC}$ & DC current density injected into active region & $4.8\times 10^2$ & $A\cdot cm^{-2}$ \\
10 & $T_0$ & characteristic temperature of the long wavelength diode & $80$ & $K$ \\
10 & $T_{0a}$ & characteristic temperature of the active region & $100$ \cite{Byrne1989A}& $K$ \\
\hline
\end{tabular}
\caption{\label{tab:example} Values of model parameters used in our numeral simulation of the rate equation. A dash (-) in the "Units" column means that the value is dimensionless.}
\end{table}

\section{The effects of temperature increase on several characteristics of semiconductor lasers in decoy state QKD systems}
As we stated before, Eve can increase the temperature of the semiconductor lasers by strong  light illumination\cite{Lee2017Free}. And the temperature increase will impact the output performance of semiconductor lasers. In QKD systems, any deviations from ideal models in security proofs will reduce the amount of final key. Thus, it is very important to study the effects of temperature increase on several characteristics of semiconductor lasers. Here by performing numerical simulation with the temperature dependent single mode rate equations (Eqs. (1) and (2)), we mainly study the effects of temperature increase on three characteristics which include the recovery time of the carrier density, the intensity of output pulses and the time interval between signal state and decoy state pulses.
\subsection{Recovery time of carriers density}
First, we briefly describe the variety of carrier density via time in the gain-switched mode. In the beginning, the carrier density $N(t)$ increases when large current is injected. Then when $N(t)$ reaches the temperature dependent transparent carrier density $N_0(T)$, the stimulated radiation process starts. Output laser is generated when $N(t)$ reaches the temperature dependent threshold carrier density $N_{th}(T)$. Then $N(t)$ falls quickly because of the disappearance of injected current and large amount consumptions of stimulated radiation. When $N(t)$ falls under $N_0(T)$, Eq. (1) can be simplified to $dN(t)/dt=N(t)/\tau_n(T)$, which means $N(t)$ follows the exponential decay after that. \par
Normally, $N(t)$ should return to temperature dependent initial carrier density $N_{DC}(T)$ before the next injection current comes. Otherwise, the initial carrier density of the next pulse is higher than that of the first one, which results a stronger photon pulse. The time needed for $N(t)$ to decay from $N_0(T)$ to $N_{DC}(T)$ is $t_{ed}=\tau_n(T)ln(N_0(T)/N_{DC}(T))$. Note that $N_{DC}(T)={J_{DC}\tau_n(T)}/{(qd)}$ \cite{Ohtsubo2013Semiconductor} and $N_{DC}(T)$ decreases as the temperature increases. We call the time needed for $N(t)$ to return back to $N_{DC}(T)$ the $recovery\ time$, denoted as $t_{re}$. The maximal repetition rate of the output photon pulses is 1/$t_{re}$ in practical QKD systems. $t_{ed}$ contributes most of $t_{re}$. The temperature increase leads to the elevation of $N_0(T)$ and the decrease of $N_{DC}(T)$, which prolongs $t_{ed}$ as well as $t_{re}$. To reduce the recovery time of carrier density, $J_{DC}$ should be set close to $J_{th}(T)$\cite{Nakata2017Intensity,Choi2017Critical}. However, the spontaneous emission under high $J_{DC}$ causes high dark counts, which definitely increases the quantum bit error rate and decreases the secure key rate\cite{Choi2017Critical}. So the value of $J_{DC}$ should be carefully considered in practical QKD systems. Here we set $J_{DC}=4.8\times 10^2 A\cdot cm^{-2}$ as shown in Table. \ref{tab:example}. The corresponding $N_{DC}(25^{\circ}$C)=0.3$N_{th}(25^{\circ}$C).
\begin{figure}[h]
\centering
\includegraphics[width=0.6 \linewidth]{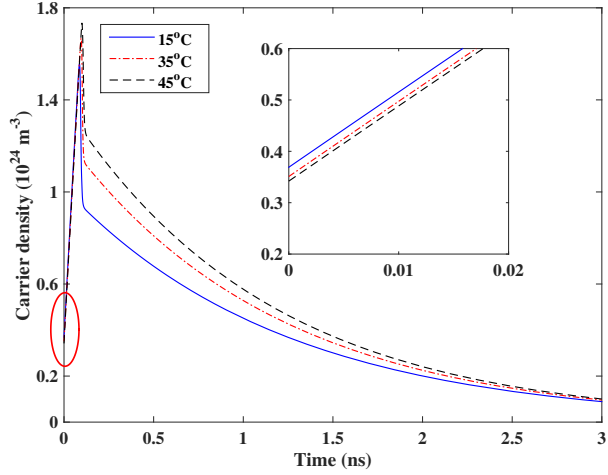}
\caption{Numerical simulation of the recovery time of carrier density at different temperatures. The solid, dash dotted and dash line represents the carrier density of semiconductor laser at $15^{\circ}$C, $35^{\circ}$C and $45^{\circ}$C respectively. Elliptic region is amplified in the inset figure. We can see that $N_{DC}(T)$ decreases as the temperature increases. $N(t)$ reaches $N_{th}(T)$ earlier for lower temperature, which leads the maximal value of $N(t)$ smaller for lower temperature. $N(t)$ follows the exponential decay after $N_0(T)$ and $N_{0}(15^{\circ}C)$ \textless$ N_{0}(35^{\circ}C)$ \textless$ N_{0}(45^{\circ}C)$.  Therefore, the lower the temperature seems to be the lower the carrier density after about 0.1ns.}
\label{fig:figure1}
\end{figure}

To simulate the variation of $N(t)$, we excite the semiconductor diode with a single current pulse. The initial carrier density is set to $N_{DC}(T)$ for different temperatures. The single injection current pulse is rectangular shape and it starts at 0$ps$ and disappears at 100$ps$. The value of the single injection current pulse is $J_{AC}$. Other parameters can be found in Table. \ref{tab:example}.\par
Fig. \ref{fig:figure1} shows the numerical simulation results of $t_{re}$ of carrier density at different temperatures. It is obvious that $t_{re}$ extends as the temperature increases. Specifically, $t_{re}=1.24ns$ at $15^{\circ}$C and $t_{re}=1.60ns$ at $45^{\circ}$C in our simulation, which leads the maximal repetition rate decreases from 806.5$MHz$ to 625.0$MHz$. In this case, the initial carrier density at $15^{\circ}$C is $N_{DC}(15^{\circ}C)=3.69\times10^{23}m^{-3}$ and the carrier density at $45^{\circ}$C is $N_{DC}(45^{\circ}C)=3.42\times10^{23}m^{-3}$. \par
If the semiconductor laser simulated above works at a frequency of 800$MHz$ and the thermoelectric cooler stabilizes the temperature at $15^{\circ}$C. $N(t)$ can always returns to $N_{DC}(15^{\circ}$C) before next current pulse comes. However, as we stated before, Eve is able to increase the temperature of the semiconductor diode to $T'$ by strong light illumination. As the temperature increases, $t_{re}$ prolongs and $N(t)$ may not return to $N_{DC}(T')$ before the next injection current applied anymore, which makes the initial carrier density of the second pulse higher than that of the first pulse. So the second photon pulse is stronger than the first one and large photon intensity fluctuation is induced\cite{Nakata2017Intensity,Choi2017Critical}.\par
\begin{figure}[h]
\centering
\includegraphics[width=0.8 \linewidth]{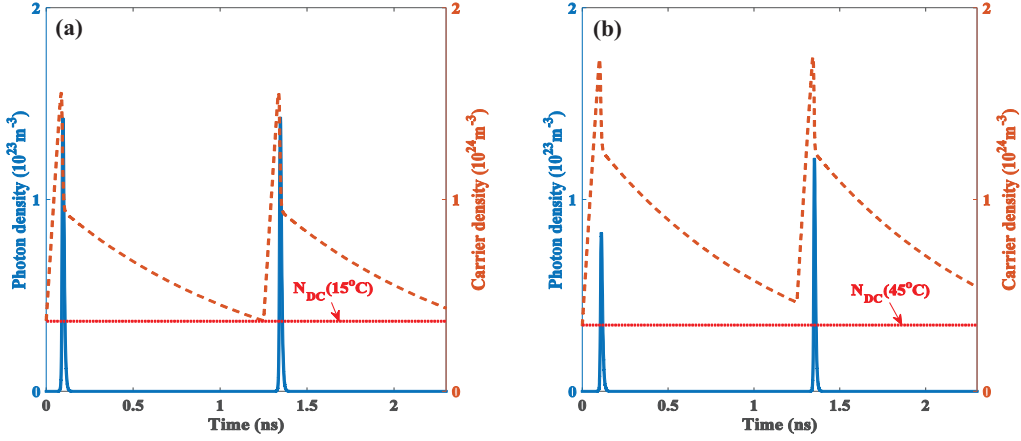}
\caption{Simulation results of Eqs. (1) and (2) at different temperatures. The solid (dash) line shows how photon (carrier) density changes with time. The dotted line represents the value of $N_{DC}(T)$. (a)Simulation results at $15^{\circ}$C. (b)Simulation results at $45^{\circ}$C.}
\label{fig:figure2}
\end{figure}

Fig. \ref{fig:figure2}(a) (Fig. \ref{fig:figure2}(b)) shows the numerical simulation results of the time-variation carrier density and photon density of the laser working at 800$MHz$ when the temperature is $15^{\circ}$C ($45^{\circ}$C). The duration of the injected current, $T_{duration}$, is also 100$ps$. The output light intensity is always assumed to be stable in QKD security proofs. However, the numerical simulation results above show that the temperature increase, which is able to be achieved by Eve using strong light illumination or microwave radiation, may strongly effect the output light intensity stability of the semiconductor lasers, especially the output light intensity stability of the lasers which work at the frequency near the maximal repetition rate. More specifically, if a laser works at a frequency of $f$ and $625MHz\textless f \leq 806.5MHz$, the laser works well at $15^{\circ}$C and the effect like the one in Fig. \ref{fig:figure2}(b) appears when temperature increases to $45^{\circ}$C. However, when $f\textless 625MHz$ which means the working frequency of the laser at $45^{\circ}$C is still lower than the maximal repetition rate of $45^{\circ}$C ($625MHz$), the effect like the one in Fig. \ref{fig:figure2}(b) will no longer appears.

\subsection{Intensity of output pulses}
The photon density reaches the maximum when $N(t)$ falls back to $N_{th}(T)$. The number of photons created by stimulated radiation is proportional to the number of carriers consumed in the active region. The initial carrier density is $N_{DC}(T)$. The carrier density injected in one cycle is ${J_{AC}T_{duration}}/{(qd)}$. So the maximal photon density is given by
\begin{center}
  \begin{equation}
    S_{max} \propto [\frac{J_{AC}T_{duration}}{qd}-N_{th}(T)+N_{DC}(T)].
   \end{equation}
\end{center}
Note that $N_{th}(T)=N_0(T)+1/[g_0(T)\Gamma \tau_p]$ and $N_{th}(T)$ increases when the temperature increases. $N_{DC}(T)$ decreases as the temperature increases. Other parameters in Eq. (9) can be treated as constant. So $S_{max}$ decreases as the temperature increases. \par
Similarly, the stimulated radiation stops when $N(t)$ drops back to $N_0{(T)}$. Therefore, the intensity of the output photon pulse is
\begin{center}
  \begin{equation}
    P_{max} \propto [\frac{J_{AC}T_{duration}}{qd}-N_0(T)+N_{DC}(T)].
   \end{equation}
\end{center}
So the output intensity also drops as the temperature increase.
\par
\begin{figure}[h]
\centering
\includegraphics[width=0.5 \linewidth]{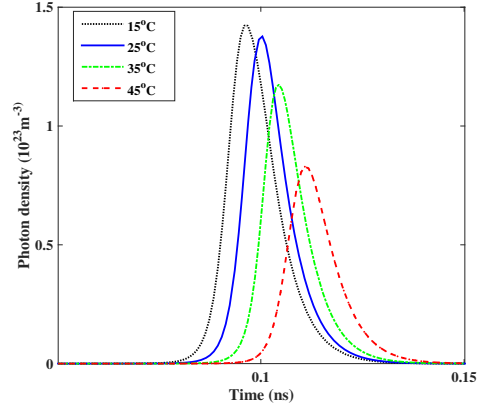}
\caption{Numerical simulation results of photon density at different temperatures. The dash, solid, dash dotted and dotted line represents the photon density of semiconductor laser at $15^{\circ}$C, $25^{\circ}$C, $35^{\circ}$C and $45^{\circ}$C respectively.}
\label{fig:figure3}
\end{figure}

Fig. \ref{fig:figure3} gives out the numerical simulation results of photon density at different temperatures. By the method of simulating the rate equations (Eqs. (1) and (2)), we can also find the fact that temperature increase leads to the decrease of $S_{max}$ (see Table. \ref{tab:table2} for detail). Similar results are founded in experiment in Ref. [\citen{Sun2015Effect}]. \par
The output light intensity is supposed to be constant in most QKD security proofs. However, our analysis and numerical simulation results show that the temperature increase in the laser decreases the output light intensity of semiconductor lasers. So Eve can break the security assumption by strong light illumination (or other tricks to heat the diode) to increase the temperature of the diode. Besides, we can also find that the time of the peak of the output photon pulse prolongs as temperature increases, which also impacts the security of practical QKD with multiple semiconductor lasers \cite{Choi2017Critical}.\par

\subsection{Time interval between signal state and decoy state pulses}
In QKD systems with "weak + vacuum" decoy state method\cite{ma2005practical}, the signal state pulses and decoy state pulses can be produced by different lasers or by the same laser. 
The decoy state pulses are excited by a lower AC injection current. In our simulation, we suppose the value of AC injection current of decoy state pulses is $2.0\times 10^{4}A\cdot cm^{-2}$ and the duration is also 100$ps$. \par
Two time dependent parameters of the semiconductor lasers should be taken into consideration in QKD systems, which includes the turn on time delay and the peak time delay. The turn on time delay is the interval from the AC injection current applied to the lasing power going out from the laser diode. Note that the laser turns on when $N(t)$ arises to $N_{th}(T)$. And the interval from the time when AC injection current applied to the time when $S(t)$ reaching maximum value is called the peak time delay. The turn on time delay of signal (decoy) state is denoted as $t_{on}(signal (decoy)\ state)$. And the peak time delay of signal (decoy) is represented by $t_{peak}(signal (decoy)\ state)$.\par

\begin{table}[h]
\centering
\begin{tabular}{cccccccc}
\hline
Temperature & $15^\text{o}C$ & $20^\text{o}C$ & $25^\text{o}C$ & $30^\text{o}C$ & $35^\text{o}C$ & $40^\text{o}C$ & $45^\text{o}C$  \\
\hline
$N_{th}(T)(10^{24}m^{-3})$ & $1.13$ & 1.16 & $1.20$ & 1.24 & 1.29 & 1.33 & 1.39 \\
$N_{DC}(T)(10^{23}m^{-3})$ & 3.69 & 3.65 & 3.60 & 3.56 & 3.51 & 3.47 & 3.42\\
$S_{max}$ of signal state $(10^{23}m^{-3})$ & 1.42 & 1.40 & 1.37 & 1.31 & 1.17 & 1.01 & 0.83 \\
$S_{max}$ of decoy state $(10^{22}m^{-3})$ & 8.82 & 7.54 & 6.33 & 4.79 & 3.26 & 2.07 & 0.57 \\
$t_{on}(signal\ state)$ $(ps)$ & $52.3$ & $56.4$ & $58.5$ & $62.1$ & $65.7$ & 69.0 & 72.9\\
$t_{peak}(signal\ state )$ $(ps)$& $95.9$ & 97.9 & 100 & $102$ & $105$ & 108 & 111\\
$t_{on}(decoy\ state)$  $(ps)$ & $63.6$ & $67.8$ & $71.3$ & $74.0$ & $80.1$&83.8&90.1 \\
$t_{peak}(decoy\ state)$ $(ps)$& $111$ & 113 & 118 & $122$ & $129$&137&156\\
\hline
\end{tabular}
\caption{\label{tab:table2} Simulation results of $S_{max}$, $t_{on}$ and $t_{peak}$ of signal state and decoy state at different temperatures. $N_{th}(T)$ is also given out.}
\end{table}
Simulation results of $S_{max}$, $t_{on}$, $t_{peak}$, $N_{DC}(T)$ and $N_{th}(T)$ at different temperatures are given out in Table. \ref{tab:table2}. As we can see, $t_{on}(signal\ state)$ ($t_{peak}(signal\ state)$) is always earlier than $t_{on}(decoy \ state)$ ($t_{peak}(decoy \ state)$). If the time of the injection current applied in the decoy state pulses is the same with that in the signals state pulses, i.e., the signal state pulses and the decoy state pulses are driven by the same clock without adjustment of individual delay, the signal state pulses are always transmitted earlier than the decoy state pulses. This phenomenon is observed in experiment in Ref. [\citen{huanganqi}]. So legitimate users should postpone the time to apply the injection current in signal state pulses, which aligns  the peaks of decoy state pulses and signal state pulses  in time frame. However, we find that the peaks of the signal state pulses and decoy state pulses are not aligned in time frame anymore when the temperature of the laser diode increases.\par
\begin{figure}[h]
\centering
\includegraphics[width=0.4\linewidth]{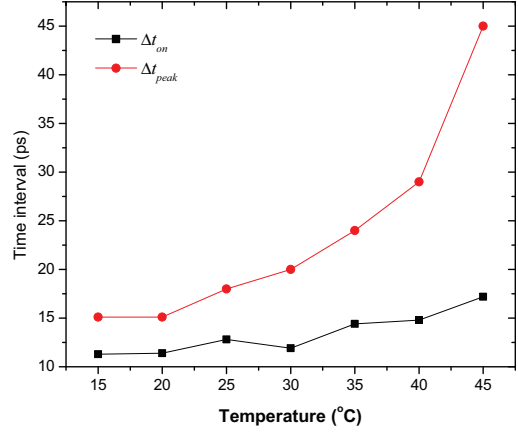}
\caption{Simulation results of $\Delta t_{on}$ and $\Delta t_{peak}$ at different temperatures. The circle line represents the the time interval between $t_{on}$ of signal state and decoy state. The square line shows the time interval between $t_{peak}$ of signal state and decoy state.}
\label{fig:figure4}
\end{figure}

Let $\Delta t_{on}=t_{on}(decoy\ state)-t_{on}(signal\ state)$ and $\Delta t_{peak}=t_{peak}(decoy\ state)-t_{peak}(signal\ state)$. Fig. \ref{fig:figure4} shows $\Delta t_{on}$ and $\Delta t_{peak}$ at different temperatures in our simulation. As we can see, $\Delta t_{on}$ varies little. However, $\Delta t_{peak}$ rises obviously as the temperature increases. Specifically, $\Delta t_{peak}$ increases from 15.1$ps$ at $15^{\circ}$C to 45.0$ps$ at $45^{\circ}$C. If the peak of signal state pulses and the peak of decoy state pulses are accurately adjusted to coincide at $15^{\circ}$C, they will mismatch with a time interval of 29.9$ps$ which is comparable with the light pulse width used in practical QKD systems. \par
The simulation results show that temperature increase in the diode of a semiconductor laser leads to a distinct raise of $\Delta t_{peak}$. Eve is able to take advantage of this temperature dependent imperfections to enlarge time dependent side channels. That is the signal state can be shifted from the decoy state by increasing the temperature. And the well-aligned signal state and decoy state pulses become distinguishable in time frame as temperature rises, which break the basic assumption of decoy state method. Then Eve can launch an improved PNS attack to steal the final key, as explained in Ref. [\citen{huanganqi}]. Moreover, the ratio of signal state intensity and decoy state intensity also rises quickly from 1.61 at $15^{\circ}$C to 14.56 at $45^{\circ}$C.
\par

\section{Quantum hacking strategies}
We have demonstrated that three main characteristics vary with temperature increase. These variations may effect the security of practical QKD systems. In this section, we will discuss quantum hacking strategies which exploit the variations of three main characteristics.
\subsection{Hacking with side channels in time dimension}
Normally, the carrier density decreases down to the steady level of $N_{DC}$ after a period of time much longer than the carrier life. If the next AC injection occurs after the carrier density becomes back to $N_{DC}$, output pulse will be identical with the previous one. However, the recovery time of carrier density increases with temperature. As shown in Fig. \ref{fig:figure2}(b), temperature increase may lead the initial carrier density of the pulse higher than that of the previous one when the semiconductor lasers are operated near the maximal repetition rate. So the intensity of the output pulse is relatively stronger than the previous one. And it takes shorter time for carrier density to reach $N_{th}(T)$ than that in the previous pulse, which leads the pulse emits earlier than the previous one. Such correlation between consecutive pulses destroys randomness whether Alice generates quantum states with true randomness or not. So Eve may take advantage of the side channels in time dimension to steal information \cite{Choi2017Critical}. For example, in a polarization based QKD system, Eve detects that one pulse emits earlier in time dimension, she knows that the same laser also fires in the previous cycle and the consecutive pulses contain the same bit information. \par
Temperature increase also decreases the output pulse intensity and prolongs the output timing of the photon pulse, which can also be exploited by Eve. For example, Eve sends strong H-polarized light to the source of a passive-basis-choice polarization based QKD system. Therefore, the temperatures of four lasers have the following relationship, $T_H$ \textgreater $T_+$ = $T_-$ \textgreater $T_V$, where $T_H$ ($T_+$, $T_-$, $T_V$) represents the temperature of laser H (+, -, V). So the output timing of four lasers are shifted, which also enlarges side channels in time dimension.\par
\subsection{Modified PNS attack on decoy state QKD systems}
As stated before, the signal state and decoy state may become distinguishable when the temperature increases. Besides, the intensity of output pulse decreases and the ratio of signal state intensity to decoy state intensity also rises as temperature increases. By combining these phenomena, we propose a modified PNS attack on decoy state QKD systems. We grant Eve with quantum nondemolition (QND) measurement ability.\par
\begin{figure}[h]
\centering
\includegraphics[width=0.25 \linewidth,angle=270]{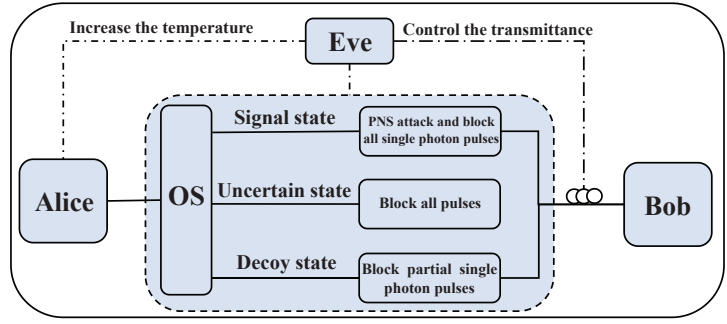}
\caption{Simple diagram of our modified PNS attack strategy. OS: optical switch.}
\label{fig:figure5}
\end{figure}
The modified PNS attack strategy is shown in Fig. \ref{fig:figure5}. First, Eve increases the temperature of semiconductor laser by some methods, such as strong light illumination. Then,  because the signal state may partially overlap the decoy state, Eve can determine the state of a photon pulse (signal state or decoy state) with a probability denoted as $p_{dis}$ by exploiting side channels in time dimension, which can be achieved by using QND measurement ability and operating the optical switch precisely. For the certain decoy state pulses, Eve blocks single photon pulses with a probability of $p_{block}$ and lets other pulses pass without disturbing. And the uncertain state pulses are all blocked by Eve. Moreover, for the certain signal state pulses, Eve mounts PNS attack and splits $n-1$ photons from $n$-photon pulses when $n\ge2$, i.e., Eve only lets one photon pass for pulses with multiple photons. And she blocks all single photon signal state pluses. To compensate the count rates of signal state and decoy state pulses, Eve also replaces the quantum channel with a lower loss one. So Eve can obtain all the information about the key. Indeed, the increase of the time interval between the signal state and decoy state pulses with temperature gives Eve the  possibility to distinguish the signal state and decoy state pulses. Besides, the decrease of the output pulse intensity and the increase of the ratio of signal state intensity and decoy state intensity provide the convenience for Eve to mount our modified PNS attack on the signal state pulses.\par
Note that no additional errors are induced in the modified PNS attack. So Eve only needs to keep the count rates of the signal state and decoy state
pulses the same with the ones without attack. Next we will theoretically show that this modified PNS attack is feasible in long transmission distance. Let us consider the QKD systems with "weak + vacuum" decoy state method. Suppose the average photon number of signal state is $\mu$ and the average photon number of decoy state is $\nu$ without attack. When the temperature increases, the average photon number of signal state (decoy state) becomes $\mu'=\alpha \mu$ ($\nu'=\beta \nu$), where 1\textgreater$\alpha$\textgreater$\beta$\textgreater0. According to Ref. [\citen{ma2005practical}], the count rate of signal state without attack is
\begin{center}
  \begin{equation}
    Q_{\mu}=\sum_{n=0}^\infty \frac{e^{-\mu}{\mu}^n}{n!}Y_n=Y_0+1-e^{-\eta\mu},
   \end{equation}
\end{center}
where $Y_n=1-(1-\eta)^n+Y_0$ is the yield of an $n$-photon state pulse, $\eta$ is the overall transmittance without attack and $Y_0$ is dark count rate. Similar, the count rate of decoy state without attack is
\begin{center}
  \begin{equation}
    Q_{\nu}=\sum_{n=0}^\infty \frac{e^{-\nu}{\nu}^n}{n!}Y_n=Y_0+1-e^{-\eta\nu}.
   \end{equation}
\end{center}
After our attack, the count rate of decoy state is
\begin{center}
  \begin{equation}
  \begin{split}
    Q_{\nu'}&=p_{dis}[\sum_{n=2}^\infty \frac{e^{-\nu'}{\nu'}^n}{n!}Y'_n+(1-p_{block})\nu'e^{-\nu'}Y'_1+(e^{-\nu'}+p_{block}\nu'e^{-\nu'})Y_0]+(1-p_{dis})Y_0\\
    &=p_{dis}(Y_0+1-e^{-\nu'\eta'}-p_{block}\nu'e^{-\nu'}\eta')+(1-p_{dis})Y_0,
   \end{split}
   \end{equation}
\end{center}
where $Y'_n=1-(1-\eta')^n+Y_0$ is the yield of an $n$-photon state pulse under attack and $\eta'$ is the overall transmittance under attack.
The count rate of signal state after our modified PNS attack is
\begin{center}
  \begin{equation}
  \begin{split}
    Q_{\mu'}&= p_{dis}[(1-\mu'e^{-\mu'}-e^{-\mu'})Y'_1+(\mu'e^{-\mu'}+e^{-\mu'})Y_0]+(1-p_{dis})Y_0.\\
    \end{split}
   \end{equation}
\end{center}
To keep the count rates of signal state and decoy state the same with those without attack, we have $Q_{\mu}=Q_{\mu'}$ and $Q_{\nu}=Q_{\nu'}$. Therefore, we can get
\begin{center}
  \begin{equation}
  \begin{split}
  \eta'=\frac{1}{1-\mu'e^{-\mu'}-e^{-\mu'}}[\frac{1}{p_{dis}}(Y_0+1-e^{-\mu\eta}-(1-p_{dis})Y_0)-(\mu'e^{-\mu'}+e^{-\mu'})Y_0]-Y_0
    \end{split}
   \end{equation}
\end{center}
\begin{center}
  \begin{equation}
  \begin{split}
   p_{block}&=\frac{1}{\nu'\eta'e^{-\nu'}}[Y_0+1-e^{-\nu'\eta'}-\frac{1}{p_{dis}}(Y_0+1-e^{-\nu\eta}-(1-p_{dis})Y_0)],
    \end{split}
   \end{equation}
\end{center}
so we can calculate $\eta'$ and $p_{block}$.\par
 The numerical simulations use some $GYS$ experiment parameters, including including the loss coefficient in the quantum channel $\delta$ = 0.21 dB/km; the dark count probability $Y_0 = 1.7 \times 10^{-6}$; the average photon number of the signal state $\mu = 0.48$; the average number of the decoy state $\nu = 0.05$; and $\eta=0.045 \times 10^{-\delta L/10}$, where $L$ is the transmittance distance. We also suppose that $\alpha=0.8$, $\beta=0.4$, $p_{dis}=0.8$ and $\eta'=0.045 \times 10^{-\delta' L/10}$, where $\delta'$ is the loss coefficient of the new replaced quantum channel and $0\leq\delta'\leq\delta$. So $\eta'\leq 0.045$. Moreover, in practice, Eve can only block a part of signal state pulses containing only one photon and $0 \textless p_{block}\textless 1$ is also needed to be satisfied.\par
\begin{figure}[h]
\centering
\includegraphics[width=0.5 \linewidth]{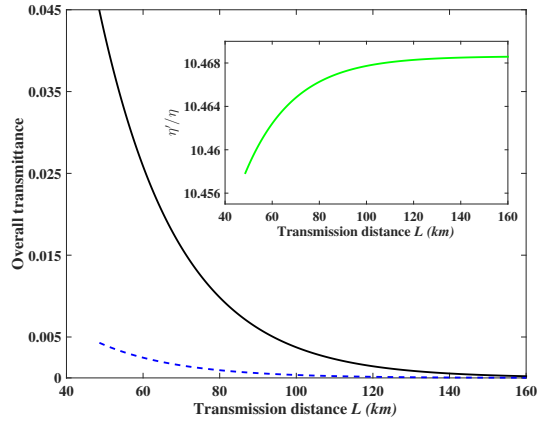}
\caption{The relationship between $\eta$ ($\eta'$) and $L$. The solid line shows how the overall transmittance $\eta'$ changes with the transmission distance $L$ after our attack and the dash line shows the relationship between the overall transmittance $\eta$ and the transmission distance $L$. The line in the inset figure represents the relationship between the ratio of $\eta'$ and $\eta$, i.e., $\eta'/\eta$, and the transmission distance $L$.}
\label{fig:figure6}
\end{figure}
Fig. \ref{fig:figure6} shows the simulation result of the relationship between $\eta$ ($\eta'$) and $L$. To keep $\eta'\leq 0.045$, we have $L \ge 48.6$km. That is to say, when the transmission distance $L$ is longer than $48.6$km, Eve can always find proper overall transmittance to keep the count rate of signal state the same with the one without attack. So we can treat $48.6$km as the secure transmission distance in our analysis. The simulation result in the inset figure shows the range of $\eta'/\eta$ and $10.456\textless \eta'/\eta\textless10.47$, which indicates that Eve is able to attack successfully by replacing the quantum channel with a lower loss one in practice. For simplicity, we use $\eta'/\eta=10$ to estimate $\delta'$. Suppose the transmission distance is $100$km and $\eta'/\eta=10^{(\delta-\delta')L/10}=10$. So we have $\delta-\delta'=0.1$dB/km and $\delta'=0.11$dB/km which may be achievable with modern technology.

\begin{figure}[h]
\centering
\includegraphics[width=0.5 \linewidth]{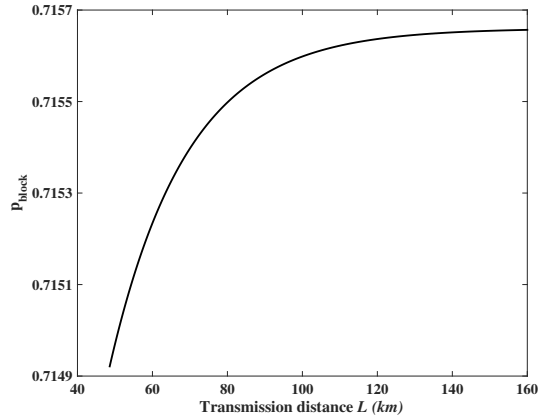}
\caption{Simulation results of the relationship between $p_{block}$ and the transmission distance $L$.}
\label{fig:figure7}
\end{figure}
Fig. \ref{fig:figure7} gives out the relationship between $p_{block}$ and $L$. We can find that $0.7149\textless p_{block}\textless 0.7157$ and $p_{block}$ almost stays invariable, which means that Eve can always keep the count rate of decoy state identical with that without attack by blocking a part of single photon decoy state pulses. All the calculation results above indicate that, by exploiting the effects of temperature increase in the semiconductor lasers, Eve can mount a modified PNS attack successfully.

\section{Conclusion and discussion}
In this paper, we study the effects of temperature increase on several key characteristics of gain-switched semiconductor lasers by performing numerical simulation on the single mode temperature dependent rate equation. The results show that temperature increase helps Eve in practical QKD systems. First, the recovery time of carrier density increases with temperature, which makes the maximal repetition rate of semiconductor lasers decrease. Second, the intensity of the output photon pulse decreases as the temperature rises. That is to say, Eve can break the security assumption of stable intensities by strong light illumination or microwave radiation. Third, the interval between the peaks of signal state and decoy state pulses increases with temperature. And Eve may distinguish signal state and decoy state pulses by raising the temperature of semiconductor lasers with proper tricks. We also propose a modified PNS attack by exploiting the effects of temperature increase in the semiconductor laser. The simulation results show that Eve can always keep the count rates of signal and decoy state the same with the ones without attack. Therefore, she can steal information about the key and stay hidden.\par
To defend such temperature dependent attacks on the semiconductor lasers, real-time monitoring on the characteristics of the output light may be helpful to detect the temperature variation in the laser. However, it is not rigorous to calculate the final key with the monitoring results on only one dimension, because temperature increase enlarges side channels in multiple dimensions, such as wavelength, time, intensity and intensity fluctuation. Besides, real-time monitoring on the intensity of the incoming light, installing isolator and band-pass filter may also help to detect or prevent Eve's strong light illumination. However, some researches showed that band-pass filter can be damaged by strong light\cite{bugge2014laser,Makarov2015Laser}. Moreover, these methods can not defeat microwave radiation from other directions in free space and other unknown tricks to increase the temperature of the semiconductor laser. So legitimate users should better monitor the temperature of the laser diodes in real time. Once the temperature increases above the threshold values, QKD systems should alarm and abort the key.\par
\section*{Acknowledgements}
This work was supported by the National Natural Science Foundation of China (61472446, 61701539 and 61501514); and the Open Project Program of the State Key Laboratory of Mathematical Engineering and Advanced Computing (2106A01).
\bibliography{sample}

\end{document}